\documentclass[epj]{svjour}

\usepackage{graphicx}
\usepackage{fancyhdr}
\def\makeheadbox{{
 }}
\def\mytitle{Dr.} 
\def\myauthors{Gudrid Moortgat-Pick}  
\def\mytype{Parallel session}
\def\mysession{Colliders - SUSY Phenomenology}

\def\mytitle{Heavy sfermions} 
\def\myauthors{Desch, Kalinowski, Moortgat-Pick, Rolbiecki, Stirling}    
\def\mytype{Contributed Talk}    
\def\mysession{Colliders - SUSY Phenomenology}

\begin{document}
\title{Heavy sfermions in SUSY analysis at LHC and ILC}
\author{K.\ Desch\inst{1}
 \and
 J.\ Kalinowski\inst{2}
\and
G. Moortgat-Pick\inst{3}\thanks{\emph{Email:} g.a.moortgat-pick@durham.ac.uk}
\and 
K. Rolbiecki\inst{2}
\and
W.J. Stirling\inst{3}
}                     
%
%
\institute{Physikalisches Institut, Universit\"at Bonn, D-53115 Bonn, Germany
\and Instytut Fizyki Teoretycznej, Uniwersytet Warszawski, PL-00681,  
Warsaw, Poland
\and
IPPP, Institute of Particle Physics Phenomenology,
University of Durham, Durham DH1 3LE, UK
}
\date{}
\abstract{The physics potential of the Large Hadron Collider in
combination with the planned International Linear Collider is discussed for
a difficult region of supersymmetry 
that is characterized by scalar SUSY particles with masses
around 2~TeV. 
Precision measurements of masses, cross
sections and forward-backward asymmetries allow to determine the
fundamental supersymmetric parameters even if only a small part of the
spectrum is accessible.  
No assumptions on a specific SUSY-breaking mechanism are imposed.
Mass contraints for the kinematically inaccessible particles can be derived. 
} 
\maketitle
%

\section{Introduction}
\label{intro}
Supersymmetry (SUSY) is one of the best-motivated candidates for physics
beyond the Standard Model (SM).
If experiments at future accelerators, the Large Hadron
Collider (LHC) and the International Linear Collider (ILC), discover SUSY they will have as well to determine precisely the 
underlying  SUSY-breaking scenario with as
few theoretical prejudices as possible.
Scenarios where the scalar SUSY
particle sector is very heavy as required, for instance, in focus-point
scenarios (FP)~\cite{focuspoint} are particularly challenging. 
In these scenarios the  gaugino masses are kept relatively
small while  squark and slepton masses might be in the multi-TeV range. 
It is
therefore of particular interest to verify whether the interplay
of a combined LHC/ILC analysis~\cite{Weiglein:2004hn} could unravel
such models with very heavy sfermions.

Methods  to derive the SUSY parameters
at collider experiments have been worked out, for instance
in~\cite{Tsukamoto:1993gt,Bechtle:2004pc}. In
\cite{Choi:1998ut,Choi:2001ww,Desch:2003vw} the
chargino and neutralino sectors have been exploited at the ILC to determine
the MSSM parameters. However, in most cases only the production processes
have been studied. Furthermore, it has been assumed that the
masses of the virtual scalar particles are already known. 
In the case of heavy scalars such assumptions, however, cannot be applied and
further observables have to be used to determine the underlying
parameters.
Studies have been made to exploit the whole production-and-decay 
process in the chargino/neutralino sector~\cite{Moortgat-Pick:1998sk}.
Exploiting spin effects, it has been shown
in~\cite{Moortgat-Pick:1999ck} that, once the chargino parameters are known,
useful indirect
bounds for the mass of the heavy virtual particles could be derived from
forward--backward asymmetries of the final lepton $A_{\rm FB}(\ell)$.

Here a FP-inspired scenario is discussed that is characterized by a
$\sim$ 2 TeV scalar particles sector~\cite{Desch:2006xp}. In addition, the neutralino
sector turns out to have very low production cross sections in
$e^+e^-$ collisions, so that it might not be fully exploitable. Only
the chargino  pair production process has high rates and
all information obtainable from this sector has to be used.
The analysis is performed entirely at the EW scale,
without any reference to the underlying SUSY-breaking mechanism.
Masses, cross sections and spin-dependent forward--backward
asymmetries are measured at the LHC and at the ILC with $\sqrt{s}\le 500$~GeV. 
The potential of a multiparameter fit to
determine the underlying parameters has been analysed. 
\section{Case study at LHC and ILC}
\label{chap2}
We study chargino production
with subsequent leptonic and hadronic decays
\begin{eqnarray}
e^{-}+e^{+} &\to& \tilde{\chi}^{+}_1+\tilde{\chi}^{-}_1,\\
\tilde{\chi}^{+}_1 &\to&
\tilde{\chi}^0_1+\ell^{+}+\nu \quad\mbox{and}\quad
\tilde{\chi}^0_1+\bar{q}_d+q_u,\label{eq:p2}\\
\tilde{\chi}^{-}_1 &\to&
\tilde{\chi}^0_1+\ell^{-}+\bar{\nu} \quad\mbox{and}\quad
\tilde{\chi}^0_1+q_d+\bar{q}_u,\label{eq:p3}
\end{eqnarray}
where $\ell=e,\mu$, $q_u=u,c$, $q_d=d,s$.
The production process
contains contributions from $\gamma$- and $Z^0$-exchange in the
$s$-channel and from $\tilde{\nu}$-exchange in the $t$-channel. The
decay processes are mediated by $W^{\pm}$, $\tilde{\ell}_{\rm L}$,
$\tilde{\nu}$ or  by $\tilde{q}_{d{\rm L}}$, $\tilde{q}_{u {\rm L}}$
exchange. The masses and eigenstates of the neutralinos and charginos are
determined by the fundamental SUSY parameters: the $U(1)$, $SU(2)$  
gaugino mass parameters
$M_1$, $M_2$, the Higgs mass parameter $\mu$ and  
the ratio of the vacuum expectation values of the two neutral Higgs fields, 
$\tan\beta=\frac{v_2}{v_1}$. 

The chosen MSSM scenario can be characterized via
the following mSUGRA parameters, taken at the GUT scale except
for $\tan\beta$:
$m_{1/2}=144~\mbox{GeV}$, $m_0=2~\mbox{TeV}$,
$A_0=0~\mbox{GeV}$, $\tan\beta=20$, $\mbox{sgn}(\mu)=+1$. 
However, our analysis is performed completely within the general
MSSM framework, without any reference to the underlying SUSY-breaking
mechanism. The parameters at the EW scale are given by
\begin{eqnarray}
M_1&=&60~\mbox{\rm GeV},\quad M_2=121~\mbox{\rm GeV},
\quad M_3=322~\mbox{\rm GeV}\nonumber\\
\mu&=&540~\mbox{\rm GeV},\quad \tan\beta=20
\end{eqnarray}
The derived masses of the SUSY
particles are listed in Table~\ref{tab:1}.
As can be seen, the charginos and neutralinos as well as the gluino
are rather light, whereas the scalar SUSY particles have masses
about 2~TeV.

\subsection{Expectations at the LHC}\label{lhc}
All
squarks in this scenario
are kinematically accessible at the LHC. The largest squark
production cross  section is for $\tilde{t}_{1,2}$. However, with
stops decaying mainly to $\tilde{g}t$ (with $BR(\tilde{t}_{1,2}\to
\tilde{g} t)\sim 66\%$), where background from top production will
be large, no new interesting channels are open in their decays.
The other squarks decay mainly via $\tilde{g} q$, but since the squarks
are very heavy, $m_{\tilde{q}_{\rm L,R}}\sim 2$~TeV, precise mass
reconstruction will be difficult.

Since the gluino is rather light in this scenario, 
several gluino decay channels can be exploited. The
largest branching ratio for the gluino decay in our scenario is 
a three-body  decay  into
neutralinos, $BR(\tilde{g}\to \tilde{\chi}^0_2 b
\bar{b})\sim 14\%$, 
followed by a subsequent three-body leptonic neutralino decay
$BR(\tilde{\chi}^0_2\to 
\tilde{\chi}^0_1 \ell^+ \ell^-)$, $\ell=e,\mu$ of about 6\%, see
Table~\ref{tab:2}. In this channel the dilepton edge will be clearly
visible, since this process has low
backgrounds~\cite{Weiglein:2004hn}. The mass difference between the
two light neutralino masses can be measured from the dilepton edge
with an uncertainty of about \cite{Kawagoe:2004rz}:
\begin{equation}
\delta(m_{\tilde{\chi}^0_2}-m_{\tilde{\chi}^0_1})\sim 0.5~\mathrm{ GeV}.
\label{eq-massdiff}
\end{equation}
The gluino mass can be reconstructed in a manner similar to   
 the one proposed in \cite{Gjelsten:2005aw}, where the SPS1a scenario is
 analysed.  
 Although our scenario is different, in both we are systematics limited due to
 hadronic energy scale and a similar relative uncertainty of $\sim$2\% can be
 expected.
\subsection{Expectations at the ILC}\label{ilc} 
At the first stage of the ILC, 
$\sqrt{s}\le 500$~GeV, only light charginos and
neutralinos are kinematically accessible. However, in this scenario
the neutralino sector is characterized by very low production cross
sections, below 1~fb, so that it might not be fully exploitable.
The low cross sections are due to the mixing character of the 
neutralinos, see~\cite{Desch:2006xp}.

Only the chargino pair production process has high rates.
We constrain our analysis to the first stage of the ILC with
$\sqrt{s} \le 500$~GeV and study only the
$\tilde{\chi}^+_1\tilde{\chi}^-_1$ production.

The chargino mass can be measured at $\sqrt{s}=350$ and $500$~GeV in
the continuum, with an error of about 
$ 0.5$~GeV~\cite{Martyn:1999tc,Aguilar-Saavedra:2001rg}.
 This can serve to
optimize the ILC scan at the threshold which,
because of the steepness of the $s$-wave excitation curve in
$\tilde{\chi}^+_1\tilde{\chi}^-_1$ production, can be used to
determine the light chargino mass very precisely, to
about~\cite{Aguilar-Saavedra:2001rg}:
\begin{equation}
m_{\tilde{\chi}^{\pm}_1}=117.1 \pm 0.1 ~\mathrm{ GeV}.
\label{eq_massthres}
\end{equation}
The mass of the lightest neutralino
$m_{\tilde{\chi}^0_1}$ can be derived via the decays of the light chargino,
either from the energy
distribution of the lepton $\ell^-$ ($BR(\tilde{\chi}^{-}_1\to
\tilde{\chi}^0_1 \ell^-
\bar{\nu}_\ell)\sim 11\%$, see Table~\ref{tab:2})
or from the invariant mass distribution of the two jets
in  hadronic decays ($BR(\tilde{\chi}^-_1\to \tilde{\chi}^0_1 q_d
\bar{q}_u)\sim 33\%$, see Table~\ref{tab:2}). 
We  take~\cite{Aguilar-Saavedra:2001rg}  
\begin{equation}
m_{\tilde{\chi}^{0}_1}=59.2 \pm 0.2~\mathrm{ GeV}.
\label{eq_masslsp}
\end{equation}
Together with the information from the LHC, Eq.~(\ref{eq-massdiff}),
a mass uncertainty for the second lightest neutralino of about
\begin{equation}
m_{\tilde{\chi}^{0}_2}=117.1 \pm 0.5 ~\mathrm{ GeV}
\label{eq_masschi02}
\end{equation}
can be assumed.

We identify the chargino pair
production process, $e^+e^- \to \tilde{\chi}_1^+ \tilde{\chi}_1^-$,
in the fully leptonic ($\ell^+\nu\tilde{\chi}^0_1\ell^-
\bar{\nu}\tilde{\chi}^0_1$) and semileptonic
($\ell\nu\tilde{\chi}^0_1q\bar{q}'\tilde{\chi}^0_1$) final states
(where $\ell=e,\mu$). We estimate an overall selection efficiency of
50\%. For both final states, $W^+W^-$ production is expected to be
the dominant SM background. For the
semileptonic (slc) final state, this background can be efficiently
reduced from the reconstruction of the hadronic invariant mass. 
 
In Table~\ref{tab:3}, we list cross sections multiplied by the
branching fraction $B_{slc}=2\times BR(\tilde{\chi}^+_1 \to
\tilde{\chi}^0_1 \bar{q}_d q_u)\times BR(\tilde{\chi}^-_1 \to
\tilde{\chi}^0_1 \ell^- \bar{\nu})+ [BR(\tilde{\chi}^-_1 \to
\tilde{\chi}^0_1 \ell^- \bar{\nu})]^2\sim 0.34$ (first two families) including an $e_{slc}=50\%$ selection efficiency.
 The error includes the statistical uncertainty based on 
${\cal L}=200$~fb$^{-1}$ in each polarization
configuration, $(P_{e^-},P_{e^+})=(-90\%,+60\%)$ and
$(+90\%,-60\%)$, and a relative uncertainty in the polarization of
$\Delta P_{e^\pm}/P_{e^\pm}=0.5\%$~\cite{Moortgat-Pick:2005cw}.

The statistical error on the forward-backward asymmetry $A_{\rm FB}$, based on binomial distribution,
is given by 
\begin{equation}
\Delta(A_{\rm FB})^{\mathrm{stat}}=2 \sqrt{\epsilon (1-\epsilon)/N},
\label{error-afb}
\end{equation}
where $\sigma_{\rm F,B}$ are the acceptance-corrected cross sections,  
$\epsilon=\sigma_{\rm F}/(\sigma_{\rm F}+\sigma_{\rm B})$ and $N$
denotes the 
number of selected events.
In Table~\ref{tab:3} the asymmetries are
listed only for the $(P_{e^-},P_{e^+})=(-90\%,+60\%)$ case, since the cross
sections for 
the opposite polarization are very small and the
statistical errors become
very large. Consequently we do not include them in the following analysis.

\section{Parameter determination}
\label{chap3}
We determine the underlying SUSY parameters in several steps:\\
$\bullet$ In the first step we use only
the masses of $\tilde{\chi}^{\pm}_1$, $\tilde{\chi}^0_1$,
$\tilde{\chi}^0_2$ and the chargino pair production cross section,
including the full leptonic and the semileptonic decays as
observables. A four-parameter fit for the parameters $M_1$, $M_2$,
$\mu$ and $m_{\tilde{\nu}}$ has been applied.\\
$\bullet$ 
In the second step we include as an additional observable the leptonic
  forward--backward
asymmetry. Only the semileptonic and purely leptonic decays were
used. The $SU(2)$ relation between the two virtual masses
$m_{\tilde{\nu}}$ and
  $m_{\tilde{e}_{\rm L}}$ has
been applied as an external constraint.\\
$\bullet$ 
As an attempt to test the $SU(2)$ mass relation for the slepton and sneutrino
  masses, in the last step both the leptonic and hadronic forward--backward
  asymmetries have been used.
A six-parameter fit for the parameters $M_1$, $M_2$, $\mu$, $m_{\tilde{\nu}}$,
$m_{\tilde{e}_{\rm L}}$ and $\tan\beta$
has been  applied.\\

\noindent {\bf a) Analysis without forward--backward asymmetry}

\noindent We use as observables the masses $m_{\tilde{\chi}^{\pm}_1}$,
$m_{\tilde{\chi}^0_{1,2}}$
 and the polarized chargino cross section multiplied by
the branching ratios of semileptonic chargino decays, see Section~\ref{ilc} 
and Table~\ref{tab:3}.

\begin{figure}
\includegraphics[width=.49\textwidth,height=0.16\textwidth,angle=0]{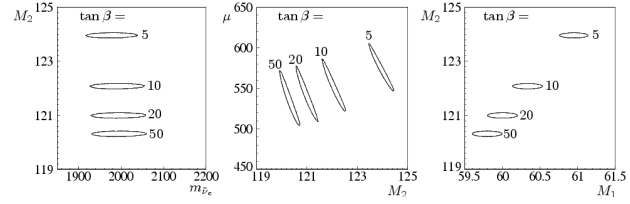}
\caption{Migration of 1$\sigma$ contours for $\tan\beta=5,\,
10,\, 20,\, 50$ with the other two parameters fixed at the values
determined by the minimum of $\chi^2$ for each 
$\tan\beta$~\cite{Desch:2006xp}.}
\label{fig:1}       
\end{figure}

We apply a four-parameter fit for the parameters $M_1$, $M_2$,
$\mu$ and $m_{\tilde{\nu}_e}$ for fixed values of $\tan\beta=5$, 10, 15, 20,
25, 30, 50 and 100. Fixing $\tan\beta$ is necessary for a
proper convergence of the {\it fitting}  
procedure because of the
strong correlation among parameters~\cite{Desch:2006xp}. We perform a
$\chi^2$ test.
It turns out that  for $\tan\beta<1.7$
the measurements are inconsistent with theoretical predictions at
least at the 1$\sigma$ level. We obtain:
\begin{eqnarray*}
  59.4  \le& M_1 & \le 62.2~\mathrm{ GeV},\\ 
 118.7  \le& M_2& \le 127.5~\mathrm{ GeV},\\ 
 450  \le& \mu& \le 750~\mathrm{ GeV},\\
1800  \le& m_{\tilde{\nu}_e}&\le 2210~\mathrm{ GeV}.
\end{eqnarray*}
Figure~\ref{fig:1} shows the
migration of 1$\sigma$ contours in $m_{\tilde{\nu}_e}$--$M_2$
(left), $M_2$--$\mu$ (middle) and $M_1$--$M_2$ (right), the
other two parameters being fixed at the values determined by the minimum
of $\chi^2$ for $\tan\beta$ changing from 5 to 10, 20 and 50.\\

\noindent {\bf b) Analysis including leptonic forward--\\backward asymmetry}

\noindent We now extend the fit by using as
additional observable the leptonic forward--backward asymmetry for
polarized beams $(-90\%,+60\%)$.

 As a result the multiparameter 
fit strongly improves the results. No assumption on
$\tan\beta$ has to be made in the fit since for too small or too large 
values of $\tan\beta$ the wrong value of $A_{\rm FB}$ is predicted.
 We find
\begin{eqnarray*}
59.7\le &M_1&\le 60.35~\mathrm{ GeV},\\
119.9\le &M_2& \le 122.0~\mathrm{ GeV},\\
500\le &\mu& \le 610~\mathrm{ GeV},\\ 
1900\le &m_{\tilde{\nu}_e}& \le 2100~\mathrm{ GeV},\\
14\le &
\tan\beta & \le 31. \label{eq:leptonic}
\end{eqnarray*}
 The constraints for the mass $m_{\tilde{\nu}_e}$
are improved by a factor of about $2$ and for gaugino mass
parameters $M_1$ and $M_2$ by a factor of about $5$, as compared to the
results of previous section with unconstrained
$\tan\beta$. The error for the higgsino mass parameter $\mu$ also
decreases significantly.
Taking the constraints of eq.~(\ref{eq:leptonic}) leads to a prediction of
\begin{eqnarray*}
&& 506 < m_{\tilde{\chi}^0_3} < 615 \mbox{GeV}, \label{eq_mchi3_pred}\\ 
&& 512 < m_{\tilde{\chi}^0_4} < 619 \mbox{GeV}, \label{eq_mchi4_pred}\\
&& 514 < m_{\tilde{\chi}^{\pm}_2} < 621\mbox{GeV}. \label{eq_mchi2_pred}
\end{eqnarray*}

\noindent {\bf c) Analysis including  hadronic and leptonic
forward--backward asymmetries: test of {\boldmath $SU(2)$}}

\noindent With the constraints for the
squark masses from the LHC, the hadronic forward--backward asymmetry
could be used to control the sneutrino mass. The leptonic
forward--backward asymmetry provides constraints on the selectron
mass and the SU(2) relation could be tested.

We perform a scan of the parameter space, 
taking light chargino and neutralino masses, 4 cross section and leptonic
 asymmetry 
measurements
and apply a $\chi^2$ test.
We derive the  following constraints:
\begin{eqnarray*}
59.30\le &M_1&\le 60.80~\mathrm{ GeV},\\
117.8\le &M_2&\le 124.2~\mathrm{ GeV},\\ 
420\le &\mu&\le 950~\mathrm{ GeV},\\ 
1860\le &m_{\tilde{\nu}_e}&\le 2200~\mathrm{ GeV},\\
 1400~\mathrm{GeV}\le &m_{\tilde{e}_{\rm L}},&\\ 
11\le& \tan\beta.&
\label{eq:nosu2}
\end{eqnarray*}
Including hadronic forward--backward asymmetry improves
the constraints as follows:
\begin{eqnarray*}
59.45\le &M_1&\le 60.80~\mathrm{ GeV},\\
 118.6\le &M_2&\le 124.2~\mathrm{ GeV},\\ 
420\le &\mu&\le 770~\mathrm{ GeV},\\
1900\le &m_{\tilde{\nu}_e} &\le 2120~\mathrm{ GeV},\\
1500~\mathrm{GeV}\le& m_{\tilde{e}_{\rm L}},&\\
11\le & \tan\beta & \le 60. \label{eq:hadnosu2}
\end{eqnarray*}
The most significant change is for the sneutrino mass, for which error bars
become smaller by $\sim 50\%$. Also an upper limit on $\tan\beta$ is found.
However we do not get an upper limit for the selectron mass.
Nevertheless, the results  for the
selectron and sneutrino masses are consistent with the $SU(2)$ relation.

\section{Conclusions}
\label{chap4}
Scenarios with heavy scalar particles 
seem to  be very challenging  for determining the MSSM parameters 
since only a very limited amount of
experimental information can be accessible.

A very powerful tool in this kind of analysis
turns out to be the forward--backward asymmetry.
 This asymmetry is strongly dependent on
the mass of the exchanged heavy particle. If the $SU(2)$ constraint
is applied, the slepton masses can be determined to a precision of
about 5\% for masses around 2~TeV at the ILC running at 500~GeV.
Although only very limited information is available,  
powerful predictions for the heavier
charginos / neutralinos can be made. \\

\noindent {\bf Acknowledgements}\\
This work was supported in part by the Polish Ministry of Science and Higher Education Grant
1 P03B10830 and by the European Community's Marie-Curie Research Training Network MRTN-CT-2006-035505 (HEPTOOLS).
%
\begin{table}
\caption{Masses of the SUSY particles (in GeV).
\label{tab:1} }  
\begin{tabular}{ccccccc}
\hline\noalign{\smallskip}
 $m_{\tilde{\chi}^{\pm}_1}$
& $m_{\tilde{\chi}^{\pm}_2}$ & $m_{\tilde{\chi}^{0}_1}$  &
$m_{\tilde{\chi}^{0}_2}$ & $m_{\tilde{\chi}^{0}_3}$ & $m_{\tilde{\chi}^{0}_4}$ &
$m_{h}$\\\noalign{\smallskip}\hline\noalign{\smallskip}
117 & 552 & 59 & 117 & 545 & 550 & 119\\ \hline\hline
$m_{H,A}$ & $m_{H^{\pm}}$ & $m_{\tilde{\nu}_e}$ &
$m_{\tilde{e}_{\rm R}}$ & $m_{\tilde{e}_{\rm L}}$ & $m_{\tilde{\tau}_1}$ &
$m_{\tilde{\tau}_2}$\\\noalign{\smallskip}\hline\noalign{\smallskip}
 1934 & 1935 & 1994 & 1996 & 1998 & 1930 & 1963 \\\hline\hline
$m_{\tilde{g}}$  &   $m_{\tilde{q}_{\rm R}}$ & $m_{\tilde{q}_{\rm L}}$ &
$m_{\tilde{t}_1}$ & $m_{\tilde{t}_2}$ & \\\noalign{\smallskip}\hline\noalign{\smallskip}
416 & 2002 & 2008 & 1093 & 1584 & & \\  \noalign{\smallskip}\hline
\end{tabular}
\end{table}

\begin{table}
\caption{Branching ratios for some important decay modes in the studied
MSSM scenario, $\ell=e,\mu,\tau$, $q_u=u,c$, $q_d=d,s$. Numbers are given
for each family separately.
\label{tab:2} }
\renewcommand{\arraystretch}{1.3}
\begin{tabular}{cccc}
\hline Mode &
$\tilde{g}\to \tilde{\chi}^0_2 b \bar{b}$ &
$\tilde{g}\to \tilde{\chi}^{-}_1 q_u \bar{q}_d$ &
$\tilde{\chi}^+_1 \to \tilde{\chi}^0_1 \bar{q}_d q_u$ \\
\hline BR &
$14.4\%$ &
$10.8\%$ &
$33.5\%$ 
\\ \hline\hline
Mode & $\tilde{\chi}^0_2\to \tilde{\chi}^0_1 \ell^+\ell^-$ &
$\tilde{t}_{1,2}\to \tilde{g} t$ &
$\tilde{\chi}^-_1 \to \tilde{\chi}^0_1 \ell^- \bar{\nu}_{\ell}$ \\
\hline BR &
$3.0\%$ &
 $66\%$ &
$11.0\%$ \\ \hline
\end{tabular}
\end{table}

\begin{table}
\renewcommand{\arraystretch}{1.3}
\caption{ Cross sections for the process $e^+e^-\to
\tilde{\chi}^+_1\tilde{\chi}^-_1$ and forward--backward asymmetries ($A_{\rm
  FB}$) 
in the leptonic $\tilde{\chi}^-_1 \to \tilde{\chi}^0_1 \ell^-
\bar{\nu}$ and hadronic $\tilde{\chi}^-_1 \to \tilde{\chi}^0_1 s
\bar{c}$ decay modes, for different beam polarization $P_{e^-}$,
$P_{e^+}$ configurations at cms $\sqrt{s}=350$~GeV and
$500$~GeV at the ILC. Concerning the errors, see text.
\label{tab:3}}
\begin{tabular}{lcc} \hline 
 & $(P_{e^-},P_{e^+})$ & $(P_{e^-},P_{e^+})$\\\hline\hline 
$\sqrt{s}=350$GeV  & $(-90\%,+60\%)$ & $(+90\%,-60\%)$\\ \hline
$\sigma(\tilde{\chi}^+_1\tilde{\chi}^-_1)$/fb &  6195.5 & 85.0   \\ 
 $\sigma(\tilde{\chi}^+_1\tilde{\chi}^-_1)\, B_{slc}\, e_{slc} $/fb  & 1062.5$\pm$4.0 & 14.6$\pm 0.7$  \\
 $A_{\rm FB}(\ell^-)$/\% & 4.42$\pm$0.29  & --  \\
 $A_{\rm FB}(\bar{c})$/\% & 4.18$\pm$0.74 & --  \\ \hline\hline
$\sqrt{s}=500$GeV & $(-90\%,+60\%)$ & $(+90\%,-60\%)$\\ \hline
 $\sigma(\tilde{\chi}^+_1\tilde{\chi}^-_1)$/fb & 3041.5 &  40.3  \\
 $\sigma(\tilde{\chi}^+_1\tilde{\chi}^-_1)\, B_{slc}\, e_{slc} $/fb  & 521.6$\pm 2.3$ &  6.9$\pm 0.4$  \\
 $A_{\rm FB}(\ell^-)$/\% & 4.62$\pm$0.41 & --  \\
 $A_{\rm FB}(\bar{c})$/\% & 4.48$\pm1.05$ & --  \\ \hline
\end{tabular}
\end{table}

\begin{thebibliography}{999}
%
\bibitem{focuspoint} 
J.~L.~Feng, K.~T.~Matchev and T.~Moroi,
Phys.\ Rev.\ D {\bf 61} (2000) 075005
[hep-ph/9909334];
J.~L.~Feng and F.~Wilczek,
hep-ph/0507032.
%
\bibitem{Weiglein:2004hn}
  G.~Weiglein et al.\  [LHC/LC Study Group],
Phys.\ Rept.\ 426 (2006) 47
  [hep-ph/0410364].
%
\bibitem{Tsukamoto:1993gt}
  T.~Tsukamoto, K.~Fujii, H.~Murayama, M.~Yamaguchi and Y.~Okada,
  Phys.\ Rev.\ D {\bf 51} (1995) 3153;
  J.~L.~Feng, M.~E.~Peskin, H.~Murayama and X.~Tata,
  Phys.\ Rev.\ D {\bf 52} (1995) 1418;
  H.~Baer et al., 
  hep-ph/9503479.

%
\bibitem{Bechtle:2004pc}
  P.~Bechtle, K.~Desch and P.~Wienemann,
  Comput.\ Phys.\ Commun.\  {\bf 174} (2006) 47 [hep-ph/0412012];
  R.~Lafaye, T.~Plehn and D.~Zerwas,
  hep-ph/0404282.
\bibitem{Choi:1998ut}
  S.~Y.~Choi, A.~Djouadi, H.~K.~Dreiner, J.~Kalinowski and P.~M.~Zerwas,
  Eur.\ Phys.\ J.\ C {\bf 7} (1999) 123;
  S.~Y.~Choi, A.~Djouadi, M.~Guchait, J.~Kalinowski, H.~S.~Song and
  P.~M.~Zerwas,
  Eur.\ Phys.\ J.\ C {\bf 14} (2000) 535;
%
J.~L.~Kneur and G.~Moultaka,
Phys.\ Rev.\ D {\bf 61} (2000) 095003.
\bibitem{Choi:2001ww}
  S.~Y.~Choi, J.~Kalinowski, G.~Moortgat-Pick and P.~M.~Zerwas,
  Eur.\ Phys.\ J.\ C {\bf 22} (2001) 563
  [Addendum, {\it ibid.\ }C {\bf 23} (2002) 769]
  [hep-ph/0108117]; S.~Y.~Choi, J.~Kalinowski, G.~Moortgat-Pick and
  P.~M.~Zerwas,
  hep-ph/0202039.
  E.~Boos, H.~U.~Martyn, G.~A.~Moortgat-Pick, M.~Sachwitz, A.~Sherstnev and P.~M.~Zerwas,
  Eur.\ Phys.\ J.\  C {\bf 30} (2003) 395
  [hep-ph/0303110];
  K.~Rolbiecki,
  Acta Phys.\ Polon.\  B {\bf 36} (2005) 3477;
  S.~Y.~Choi, B.~C.~Chung, J.~Kalinowski, Y.~G.~Kim and K.~Rolbiecki,
  Eur.\ Phys.\ J.\  C {\bf 46} (2006) 511
  [arXiv:hep-ph/0504122].
%
\bibitem{Desch:2003vw}
  K.~Desch, J.~Kalinowski, G.~Moortgat-Pick, M.~M.~Nojiri and G.~Polesello,
  JHEP {\bf 0402} (2004) 035
  [hep-ph/0312069].
%
\bibitem{Moortgat-Pick:1998sk}
G.~Moortgat-Pick, H.~Fraas, A.~Bartl and W.~Majerotto,
Eur.\ Phys.\ J.\ C {\bf 7} (1999) 113
[hep-ph/9804306];
G.~Moortgat-Pick and H.~Fraas,
Phys.\ Rev.\ D {\bf 59} (1999) 015016 [hep-ph/9708481];
G.~A.~Moortgat-Pick and H.~Fraas,
  Phys.\ Rev.\  D {\bf 59} (1999) 015016
  [hep-ph/9708481].


\bibitem{Moortgat-Pick:1999ck}
G.~Moortgat-Pick and H.~Fraas,
Acta Phys.\ Polon.\ B {\bf 30} (1999) 1999
[hep-ph/9904209];
G.~Moortgat-Pick, A.~Bartl, H.~Fraas and W.~Majerotto,
Eur.\ Phys.\ J.\ C {\bf 18} (2000) 379
[hep-ph/0007222].
%
\bibitem{Desch:2006xp}
  K.~Desch, J.~Kalinowski, G.~Moortgat-Pick, K.~Rolbiecki and W.~J.~Stirling,
  JHEP {\bf 0612} (2006) 007
  [hep-ph/0607104];
  K.~Rolbiecki, K.~Desch, J.~Kalinowski and G.~Moortgat-Pick,
  hep-ph/0605168.
%
\bibitem{Kawagoe:2004rz}
  K.~Kawagoe, M.~M.~Nojiri and G.~Polesello,
  Phys.\ Rev.\ D {\bf 71} (2005) 035008
  [hep-ph/0410160].
%
\bibitem{Gjelsten:2005aw}
  B.~K.~Gjelsten, D.~J.~Miller and P.~Osland,
  JHEP {\bf 0506} (2005) 015
  [hep-ph/0501033].

\bibitem{Martyn:1999tc}
  H.~U.~Martyn and G.~A.~Blair,
  hep-ph/9910416.


\bibitem{Aguilar-Saavedra:2001rg}
J.~A.~Aguilar-Saavedra et al.,
 hep-ph/0106315;\\
%
K.~Abe et al.,
hep-ph/0109166.
%
T.~Abe et al.,
hep-ex/0106056.
%

\bibitem{Moortgat-Pick:2005cw} 
G.~A.~Moortgat-Pick {\it et al.}, 
  hep-ph/0507011.


\end{thebibliography}
\end{document}